\documentclass[preprint]{aastex}
\usepackage{natbib}
\usepackage{amsmath,esint}
\usepackage{aasmacros}

\usepackage{bm}
\usepackage{url}

\newcommand{\obsarea}{962.3 deg$^2$}
\newcommand{\obsprob}{16.3\%}
\newcommand{\bayestar}{\textsc{BAYESTAR}}
\newcommand{\json}{\texttt{json}}
\newcommand{\lalarea}{8.0\%}

\newcommand{\fits}{\texttt{fits}}

\shorttitle{GROWTH: Real-Time DECam Search for a Counterpart to S190426c}
\shortauthors{Goldstein, Andreoni, et al.}
\begin{document}

\title{
    GROWTH on S190426c. II. Real-Time Search for a Counterpart to the Probable Neutron Star-Black Hole Merger using an Automated Difference Imaging Pipeline for DECam
}
\author[0000-0003-3461-8661]{Daniel~A.~Goldstein}
\altaffiliation{Hubble Fellow}
\affiliation{California Institute of Technology, 1200 East California Blvd, MC 249-17, Pasadena, CA 91125, USA}
\author[0000-0002-8977-1498]{Igor~Andreoni}
\affiliation{California Institute of Technology, 1200 East California Blvd, MC 249-17, Pasadena, CA 91125, USA}
\collaboration{these authors contributed equally to this work}
\author[0000-0002-3389-0586]{Peter~E.~Nugent}
\affiliation{Computational Science Department, Lawrence Berkeley National Laboratory, 1 Cyclotron Road, MS 50B-4206, Berkeley, CA 94720, USA}
\affiliation{Department of Astronomy, University of California, Berkeley, 501 Campbell Hall \#3411, Berkeley, CA, 94720, USA}

\author{Mansi~M.~Kasliwal}
\affiliation{California Institute of Technology, 1200 East California Blvd, MC 249-17, Pasadena, CA 91125, USA}

\author{Michael~W.~Coughlin}
\affiliation{California Institute of Technology, 1200 East California Blvd, MC 249-17, Pasadena, CA 91125, USA}

\author{Shreya Anand}
\affiliation{California Institute of Technology, 1200 East California Blvd, MC 249-17, Pasadena, CA 91125, USA}

\author{Joshua S. Bloom}
\affiliation{Department of Astronomy, University of California, Berkeley, 501 Campbell Hall \#3411, Berkeley, CA, 94720, USA}
\affiliation{Lawrence Berkeley National Laboratory, 1 Cyclotron Road MS 50B-4206, Berkeley, CA, 94720, USA}

\author{Jorge Mart\'inez-Palomera}
\affiliation{Department of Astronomy, University of California, Berkeley, 501 Campbell Hall \#3411, Berkeley, CA, 94720, USA}

\author{Keming Zhang}
\affiliation{Department of Astronomy, University of California, Berkeley, 501 Campbell Hall \#3411, Berkeley, CA, 94720, USA}

\author[0000-0002-2184-6430]{Tom{\'a}s Ahumada}
\affiliation{Department of Astronomy, University of Maryland, College Park, MD 20742, USA}

\author{Ashot Bagdasaryan}
\affiliation{California Institute of Technology, 1200 East California Blvd, MC 249-17, Pasadena, CA 91125, USA}

\author[0000-0001-5703-2108]{Jeff Cooke}
\affiliation{Australian Research Council Centre of Excellence for Gravitational Wave Discovery (OzGrav), Swinburne University of Technology, Hawthorn, VIC, 3122, Australia}
\affiliation{Centre for Astrophysics and Supercomputing, Swinburne University of Technology, Hawthorn, VIC, 3122, Australia}

\author{Kishalay De}
\affiliation{California Institute of Technology, 1200 East California Blvd, MC 249-17, Pasadena, CA 91125, USA}

\author{Dmitry~A. Duev}
\affiliation{California Institute of Technology, 1200 East California Blvd, MC 249-17, Pasadena, CA 91125, USA}

\author{U.~Christoffer Fremling}
\affiliation{California Institute of Technology, 1200 East California Blvd, MC 249-17, Pasadena, CA 91125, USA}

\author[0000-0002-1955-2230]{Pradip Gatkine}
\affiliation{Department of Astronomy, University of Maryland, College Park, MD 20742, USA}

\author{Matthew Graham}
\affiliation{California Institute of Technology, 1200 East California Blvd, MC 249-17, Pasadena, CA 91125, USA}

\author{Eran O. Ofek}
\affiliation{Department of Particle Physics \& Astrophysics, Weizmann Institute of Science, Rehovot 76100, Israel}

\author{Leo~P. Singer}
\affiliation{Goddard Space Flight Center, 8800 Greenbelt Rd, Greenbelt, MD 20771, USA}

\author{Lin Yan}
\affiliation{California Institute of Technology, 1200 East California Blvd, MC 249-17, Pasadena, CA 91125, USA}

\begin{abstract}
The discovery of a transient kilonova following the gravitational-wave event GW\,170817 highlighted the critical need for coordinated rapid and wide-field observations, inference, and follow-up across the electromagnetic spectrum. In the Southern hemisphere, the Dark Energy Camera (DECam) on the Blanco 4-m telescope is well-suited to this task, as it is able to cover wide-fields quickly while still achieving the depths required to find kilonovae like the one accompanying GW170817 to $\sim$500  Mpc, the binary neutron star  horizon distance for current generation of LIGO/Virgo collaboration (LVC) interferometers. Here, as part of the multi-facility followup by the Global Relay of Observatories Watching Transients Happen (GROWTH) collaboration,  we describe the observations and automated data movement, data reduction, candidate discovery, and vetting pipeline of our target-of-opportunity DECam observations of S190426c, the first possible neutron star--black hole merger detected via gravitational waves. Starting 7.5hr after S190426c, over 11.28\,hr of observations, we imaged an area of 525\,deg$^2$ ($r$-band) and 437\,deg$^2$ ($z$-band); this was 16.3\% of the total original localization probability and nearly all of the probability density visible from the Southern hemisphere. The machine-learning based pipeline was optimized for fast turnaround, delivering transient candidates for human vetting within 17 minutes, on average, of shutter closure. We reported nine promising counterpart candidates 2.5 hours before the end of our observations. Our observations yielded no detection of a bona fide counterpart to $m_z = 22.5$ and $m_r = 22.9$ at the 5$\sigma$ level of significance, consistent with the refined LVC positioning. We view these observations and rapid inferencing as an important real-world test for this novel end-to-end wide-field pipeline.
\end{abstract}

\section{Introduction}
\label{sec:intro}

Joint detections of electromagnetic (EM) and gravitational waves (GWs) from compact binary mergers involving neutron stars (NSs) are a promising new way to address a number of open questions in astrophysics and cosmology \citep[see, e.g.,][for reviews]{2009arXiv0902.1527B,2019arXiv190402718C}.
The combined EM/GW dataset from the binary neutron star (BNS) merger GW170817 \citep{mma} provided  a high-precision measurement of the speed of gravity \citep{speedgrav}, gave new insight into the origin of the heavy elements \citep[e.g.,][]{2017Natur.551...80K,2017Natur.551...67P,2017ApJ...848L..19C,2017Sci...358.1570D,2017Natur.551...75S,2017Sci...358.1565E,2017Sci...358.1556C,2018ApJ...855...99C,2018arXiv181000098S,2019PhRvL.122f2701W,2019MNRAS.tmpL..14K,2019arXiv190501814J,2019ApJ...875..106C}, demonstrated a novel technique for measuring cosmological parameters \citep{2017Natur.551...85A}, and provided unparalleled insight into the radiation hydrodynamics of compact binary mergers \citep[e.g.,][]{2017ApJ...848L..20M,2017Sci...358.1559K,2017Sci...358.1579H,2017ApJ...848L..21A,2018Natur.554..207M,2019Sci...363..968G,2018PhRvL.120x1103L}.
To date, GW170817 remains the only astrophysical event that has been detected in both the EM and GW messengers.
To realize the full scientific potential of BNS and NS-black hole (BH) mergers with joint EM/GW detections, many more  must be discovered and followed up.

The current working procedure for joint EM/GW astronomy begins when a network of GW observatories \citep[presently  LIGO, the  Laser Interferometer Gravitational-Wave Observatory,  and  the Virgo Gravitational-Wave Observatory;][]{2015CQGra..32g4001L,2015CQGra..32b4001A} detects a GW source, and, by analyzing its waveform, localizes it to a region of the sky that is typically between 100 and 1000 deg$^2$. Nearly contemporaneous $\gamma$-rays and X-rays may be detected and localized if the merger also produces a short gamma-ray burst (GRB) at a favorable viewing angle \citep[see, e.g.,][]{1989Natur.340..126E,2006ApJ...638..354B}.
It then falls to the optical and near-infrared observational communities to search for transient events  in the large localization region that are consistent with theoretical expectations for spectrum synthesis in compact binary mergers, enabling the GW sources to be localized precisely (i.e., associated with a host galaxy).
Such transients, often referred to as ``kilonovae'' because they are roughly $10^3$ times brighter than novae, are powered by the rapid decay of $r$-process material synthesized in the mergers \citep{2010MNRAS.406.2650M}, and they are distinguished from other transients by their rapidly evolving light curves, which fade and redden in just a few days \citep[e.g.,][]{2013ApJ...775..113T,2013ApJ...775...18B}.
In order to search  large areas of sky for such faint and rapidly evolving transients, telescopes with large apertures, imagers with large fields of view, and pipelines that can rapidly process images  to efficiently identify transient candidates are required.

In the Southern Hemisphere, the Dark Energy Camera \citep[DECam;][]{2015AJ....150..150F} on the Victor M. Blanco 4-meter Telescope at  Cerro Tololo Inter-American Observatory (CTIO) is a  powerful instrument for detecting kilonovae associated with gravitational wave triggers.
The wide field of view ($\sim$3 deg$^2$) of the instrument, combined with its red sensitivity and the substantial aperture of its telescope, make it well suited to follow up even the most distant BNS and NS-BH mergers in the LIGO/Virgo horizon.
The power of DECam for EM/GW follow-up was illustrated by its significant role in the study of AT2017gfo, the kilonova associated with GW170817 \citep{2017ApJ...848L..16S, 2017ApJ...848L..17C}, and by its important role in the follow-up of several other GW events from LIGO and Virgo \citep{2016ApJ...823L..33S,2016ApJ...826L..29C,2016ApJ...823L..34A,2019ApJ...873L..24D}.

In preparation for the third LIGO/Virgo GW observing run (O3), we developed a high-performance image subtraction pipeline to rapidly identify transients on DECam images.
The National Optical Astronomy Observatory (NOAO), which allocates time on DECam, granted our team the opportunity to trigger the instrument to follow up neutron star mergers detected in gravitational waves by LIGO and Virgo during the first half of O3 (NOAO Proposal ID 2019A-0205; PIs Goldstein and Andreoni).
We activated our first trigger on the unusual GW source S190426c \citep{gcn2}, potentially the first neutron star-black hole (NS-BH) merger to be detected by LIGO and Virgo.
In this Letter, we describe our follow-up observations of this event, with a focus on the software infrastructure we have developed to rapidly conduct wide-field optical follow-up observations of neutron star mergers using DECam.

\section{S190426c: A Probable NS-BH Merger}
\label{sec:event}

On 2019 April 26 at  15:21:55 UTC, the LIGO Scientific Collaboration and Virgo Collaboration (LVC) identified a compact binary merger candidate, dubbed ``S190426c,'' during real-time processing of data from LIGO Hanford Observatory, LIGO Livingston Observatory, and Virgo Observatory.
The candidate was detected by four separate analysis pipelines: GstLAL \citep{2017PhRvD..95d2001M}, MBTAOnline \citep{2016CQGra..33q5012A}, PyCBC Live \citep{2017ApJ...849..118N}, and SPIIR, with a false alarm rate of 1 in 1.7 years.
Roughly twenty minutes after detecting the event, LVC issued a circular on the NASA Gamma-Ray Coordinates Network (GCN)\footnote{\url{https://gcn.gsfc.nasa.gov/gcn3_archive.html}} reporting the discovery \citep{gcn1}.

\begin{figure*}[!htbp]
    \centering
    \includegraphics[width=1\textwidth]{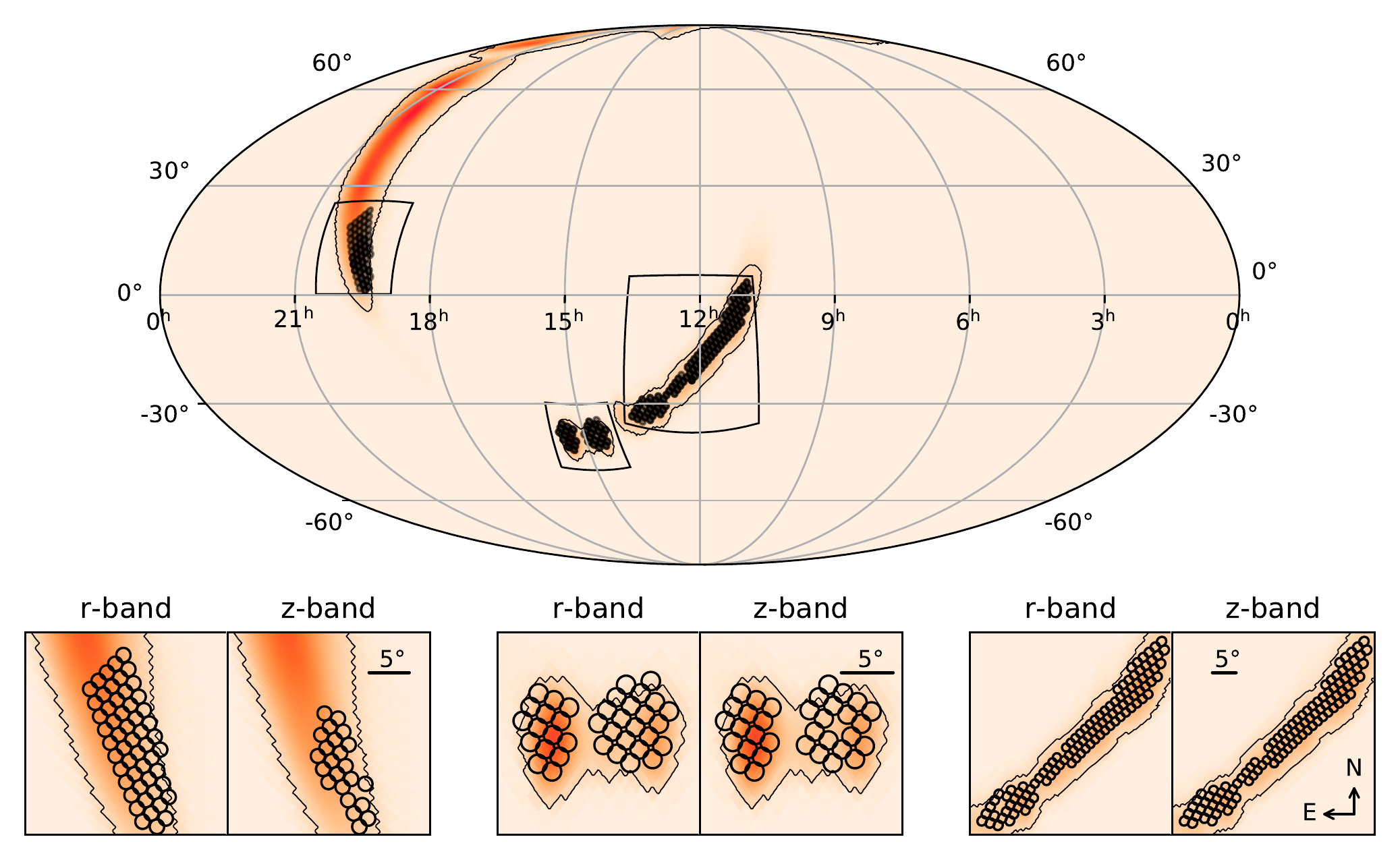}
    \caption{DECam sky coverage overplotted on the initial \bayestar\ skymap.
    Each black circle represents one DECam pointing, where the 3 deg$^2$ DECam field of view has been approximated as a circle of radius 0.9 deg.
    Solid black lines show contours of the \bayestar\ 90\% confidence region.
    Nearly uniform coverage of the region of the skymap visible from CTIO was obtained in the $r$ and $z$ bands.
    In total, \obsprob\ of the \bayestar\ probability was enclosed by the observations.
    This probability dropped to \lalarea\ when the refined LALInference skymap (not shown) was released.}
    \label{fig:skymap}
\end{figure*}

The initial GCN included a preliminary skymap giving a probabilistic localization  of the event from the  \bayestar\ rapid GW localization code \citep[see Figure \ref{fig:skymap}]{2016PhRvD..93b4013S}.
The total area of sky covered by the 90\% confidence region was 1262 deg$^2$, with an estimated luminosity distance of 375 $\pm$ 108 Mpc.
As Figure \ref{fig:skymap} shows, the probability was concentrated in two distinct regions on the sky, one largely north of the celestial equator at $\mathrm{RA}\approx20.5$h, and another region south of the equator roughly centered at $\mathrm{RA}\approx13.5$h.

The initial classification of the event was consistent with several possible progenitor scenarios.
The initial GCN circular classified the event as a BNS merger with a probability of 49\%, a compact binary merger with at least one object with a mass in the hypothetical “mass gap” between neutron stars and black holes  (3--5 solar masses) with a probability of 24\%, a terrestrial event (ie., not astrophysical) with a probability of 14\%, and a neutron star-black hole (NS-BH) merger with a probability of 13\% \citep{gcn2}.
These probabilities were later updated in favor of the NS-BH interpretation, which was assigned a revised probability of 73.1\% (including the mass gap probability), with no change to the probability of being a terrestrial event \citep{gcn3}.
Given the significant probability of the event originating from a NS merger, we decided to trigger our DECam program to search for an optical counterpart.

\section{Observations}

We triggered DECam follow-up of S190426c under NOAO proposal 2019A-0205 (PIs Goldstein \& Andreoni), publishing a GCN circular describing our plan for the observations \citep{gcn4} and our intentions to make the data public immediately. We adopted an integrated observing strategy using the $r$ and $z$ filters with 30\,s and 50\,s exposures, respectively.  The visits in $r$ and $z$ were spaced in time by at least 30 minutes to facilitate the rejection of moving objects.  We observed from 2019-04-26 22:57:35 until 2019-04-27 10:25:54 UT, for a total of 11.28\,hr.  We acquired 196 exposures in $r$ and 163 in $z$, covering an area of 525\,deg$^2$ and 437\,deg$^2$ respectively, assuming an effective 60-CCD 2.68\,deg$^2$ field of view for DECam that excludes the chip gaps. 
Our observations resulted in empirical limiting magnitudes of $m_z = 22.5$ and $m_r = 22.9$ at the 5$\sigma$ level of significance.

The information provided by the GW skymap (large localization area, large distance, possible BH companion) compelled us to modify the observing strategy that we originally designed for this program, which was based on 3 visits in $g$-$z$-$g$ bands on the first night and a $g$-$z$ pair on the second night after the trigger.  Exposure times were planned to be 15\,s in $g$ and $25$\,s in $z$ band.  Such a strategy was designed to follow-up primarily BNS mergers enclosed in an error region $\lesssim 150$\,deg$^2$ in extension and $<200$\,Mpc away.  The $g-z$ filter combination is optimal to capture and recognize the rapidly-evolving blue component that BNS mergers such as GW170817 are expected to show \citep[see e.g.,][]{2017Sci...358.1565E,2017Sci...358.1574S,2019PASP..131f8004A, 2019ApJ...874...88C}. The large distance to S190426c, along with the theoretical expectation that NS-BH mergers may not show any bright blue component at early times \citep{2017Natur.551...80K}, advocated in favor of deeper exposures and redder filters.  The third visit planned for the first night was dropped in favor of a broader sky coverage with longer $z$-band exposures.
For further details on schedule optimization for our DECam program, see Andreoni, Goldstein, et al. (in preparation).

Our observations of S190426c were scheduled automatically by the GROWTH target-of-opportunity (ToO) marshal system\footnote{\url{https://github.com/growth-astro/growth-too-marshal}} described in \cite{2019PASP..131d8001C} and \cite{2019PASP..131c8003K}.
For this event, we instructed the ToO marshal to employ a ``greedy'' algorithm to generate a schedule of observations that tiled as much of the 90\%  credible position region of the initial \bayestar\ skymap as possible.
The schedule was generated before sunset in Chile on 2019 April 26 and exported as a \json\ file.
The initial \bayestar\ skymap and our series of observations are shown in Figure \ref{fig:skymap}.
The \json\ file was ingested into the DECam Survey Image System Process Integration \citep[SISPI;][]{2012SPIE.8451E..12H} readout and control system, which executed the observations.
As soon as each exposure was completed, SISPI transferred each raw exposure to  NOAO in Tucson, AZ via the Data Transport System \citep[DTS;][]{2010SPIE.7737E..1TF} for archiving.

A second epoch was planned for the following night using the same filters, but the refined skymap that  LVC released after our observations \citep{gcn7} using the more precise LALInference localization pipeline \citep{2015PhRvD..91d2003V} completely eliminated the localization probability in any sky region with DECam surveys template coverage (see Section \ref{sec:templates}), necessary to discover transients with our pipeline.  Moreover, the visible region of sky that we could have observed resides on the Galactic plane, where several magnitudes of extinction and crowded stellar fields make the detection of faint, extragalactic transients a particularly difficult task.
Therefore we decided against more disruptive ToO observations,  ending our DECam observing campaign for S190426c after a single night of data-taking.
We describe three additional discovery engines and several follow-up facilities that undertook the search for the electromagnetic counterpart to S190426c as part of the GROWTH network in a suite of companion papers (Kasliwal et al. in prep, Bhalerao et al. in prep). A synopsis of the worldwide community observations reported in GCNs can be found in \cite{2019arXiv190502186H}.

\section{Real-Time Automated Difference Imaging Pipeline}
\label{sec:pipeline}


As soon as observations commenced on the first night of our trigger, we programmatically checked the NOAO archive each second for new images from the DTS.
Each time a new image was found, we automatically downloaded it over FTP to the National Energy Research Scientific Computing Center (NERSC) in Berkeley, California and stored it on a high-performance \texttt{Lustre} parallel filesystem, making use of the ESNet energy sciences high-speed internet backbone connecting US Department of Energy  facilities.
The typical data transfer rate from Tucson to Berkeley was 40MB/s, enabling each 550 MB \texttt{fits}\ focal plane exposure to be delivered in an average transfer time of  14 seconds.

\subsection{Exposure Segmentation and Parallelization}
When each raw image arrived at NERSC, a job was programmatically launched  via  \texttt{slurm}\footnote{\url{https://slurm.schedmd.com/overview.html}}\ to process it, beginning the real-time search.
Jobs were executed on the Cray XC40 \texttt{cori} supercomputer.
Each exposure was delegated for processing to a single 64-logical core \texttt{haswell} compute node.
In each job, each of the 62 DECam science CCDs was assigned to a single logical core.
We arranged a special, low-latency ``realtime'' job queue for this project to provide near-immediate access to NERSC computer resources.
Our realtime queue gave us on-demand access to 18 \texttt{haswell} compute nodes, allowing us to process up to 18 exposures simultaneously.
We found that this allocation of computer resources was sufficient to ensure  fast turnaround.

As a first step in the processing, each raw DECam \texttt{fits} file was split into 62 separate \texttt{fits} files, one for each CCD.
Except for template generation, all subsequent pipeline steps were performed on a per-CCD  basis, using the Message Passing Interface (MPI) to facilitate the concurrent execution of 62  independent copies of the pipeline in each of up to 18 jobs running simultaneously.
The top-level pipeline code was written in the Python programming language and  run inside  a high-performance \texttt{shifter}\footnote{A docker-like containerization service for high-performance computing, see \citealt{Gerhardt_2017}.} container to increase performance on the NERSC hardware.

\subsection{Detrending and Astrometric Calibration}
The raw frames we ingested from the NOAO archive underwent no calibration, containing only \fits\ header keywords and integer pixel values, so we first performed a series of detrending and preprocessing steps to transform them into usable science frames.
For each frame, we made an overscan correction as described in \cite{2017PASP..129k4502B}.
We also generated a mask frame for each CCD, masking out any pixels above the saturation value of their amplifier.
Because our observations were time-sensitive, and because DECam is a very stable instrument, we used flat and bias frames from a previous night for the real-time processing (i.e., we did not take flats or bias frames in our observing sequence).
The flat and bias frames we used to process the data for S190426c were taken on 2018 Nov 1 as part of the DECam Legacy Survey \citep{2019AJ....157..168D}.
We subtracted the bias frames from the raw pixels  and then divided by the flat frames.
Any science pixel values rendered invalid by the flat-fielding were masked.
We applied the standard DECam bad pixel masks, but to achieve fast turnaround did not apply crosstalk corrections or correct for the brighter-fatter effect.
These effects are only relevant to high-precision photometry and have little impact on transient discovery.
We processed all science CCDs from each pointing, including those that have been deemed defective (N30 and S30).

We produced a source catalog of each detrended science image using \texttt{SExtractor} \citep{1996A&AS..117..393B} that we fed into a development-branch version of \texttt{SCAMP} \citep{2006ASPC..351..112B} to perform astrometric calibration against the \textit{Gaia}\ DR1 catalog \citep{2016A&A...595A...2G}, which consistently provided extremely reliable astrometry.

\subsection{Template Generation}
\label{sec:templates}

\begin{figure*}
    \centering
    \includegraphics[width=1\textwidth]{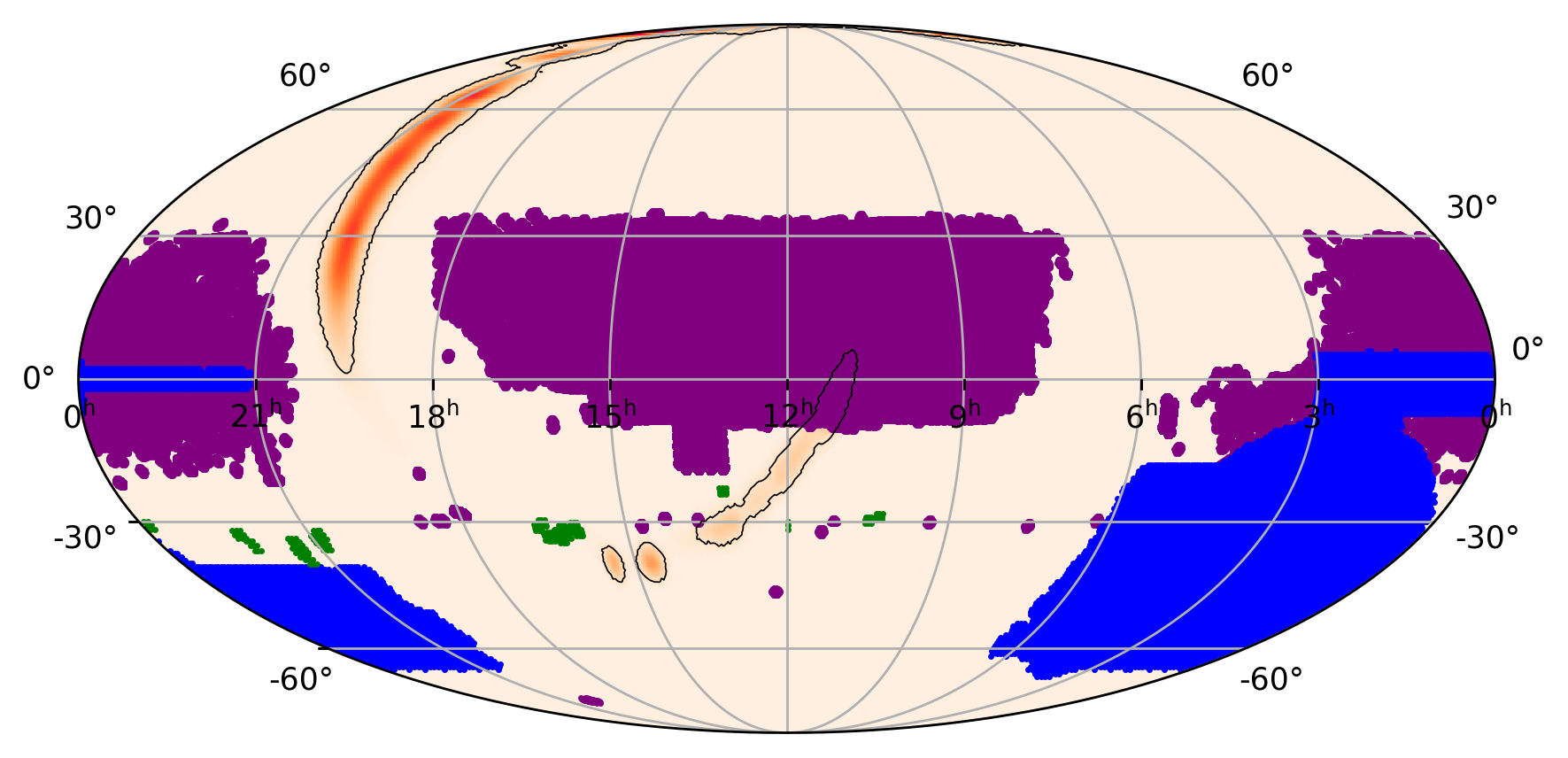}
    \includegraphics[width=1\textwidth]{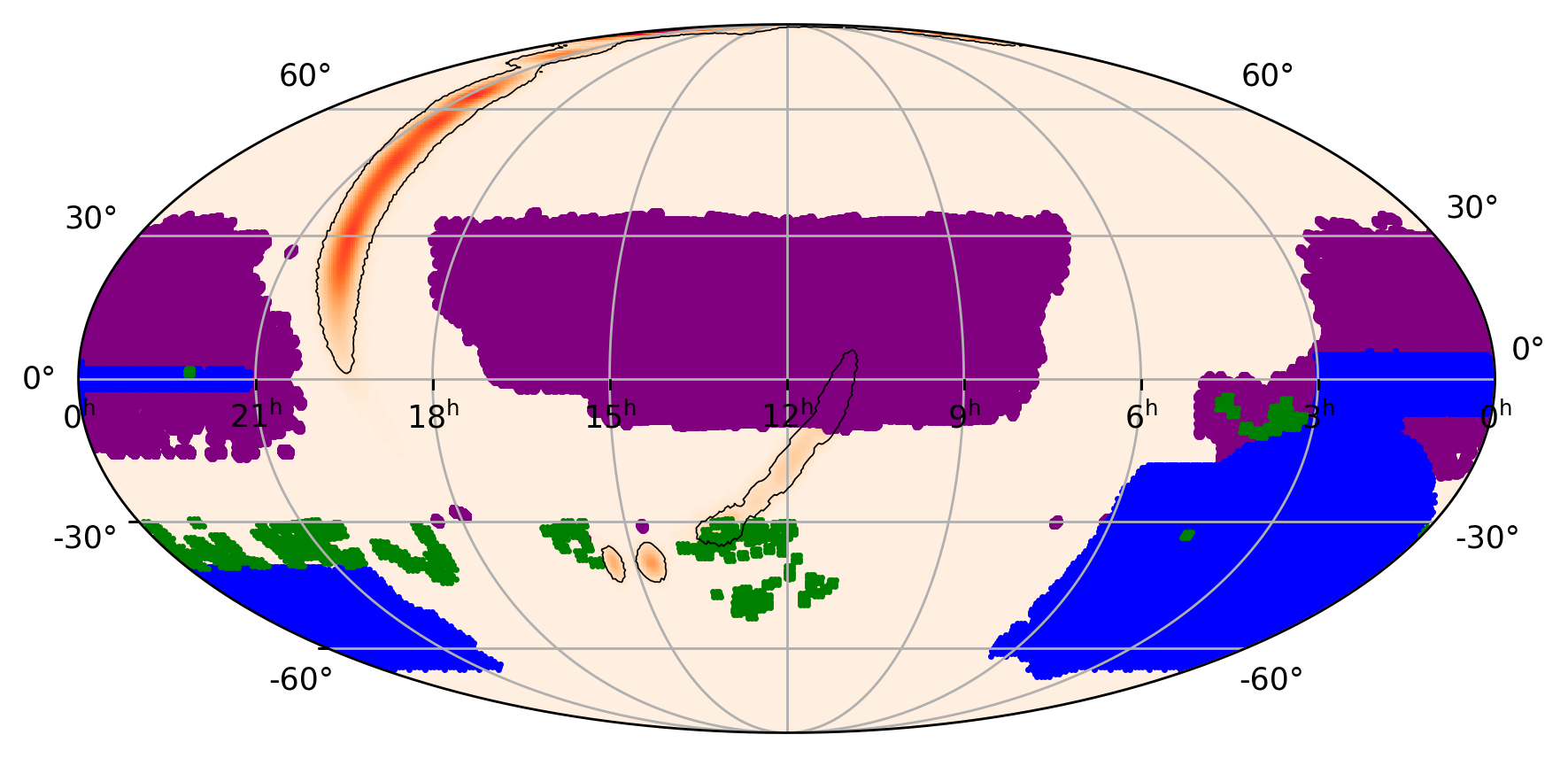}
    \caption{Accumulated $r$- (top) and $z$-band (bottom) DECam template coverage for this event. Our template bank drew from DES (blue), DECaLS (purple), and BLISS (green) images.
    }
    \label{fig:templates}
\end{figure*}

To perform image subtraction, we assembled a library of template images from three publicly available DECam datasets: the Dark Energy Survey DR1 \citep{2016MNRAS.460.1270D, 2018ApJS..239...18A}, the DECam Legacy Survey DR7 \citep{2019AJ....157..168D}, and the Blanco Imaging of the Southern Sky Survey \citep[BLISS;][]{bliss} stacked images distributed by the NOAO archive.
We downloaded all of these astrometrically and photometrically calibrated template images to disk at NERSC.
In total, the templates required about 50 TB of disk space and covered about 14,500 deg$^2$ of the sky below a declination of $+30^\circ$.
Our $r$- and $z$-band template coverage relative to the sky map of S190426c is shown in Figure \ref{fig:templates}.
Only a small region of the sky map for this event had template coverage (about 120 deg$^2$ in $z$-band, and 100 deg$^2$ in $r$- and $z$-band) from the DECaLS and BLISS surveys.
We used \texttt{SWarp} \citep{2010ascl.soft10068B} to combine and crop the individual template images into references for each CCD.
The coaddition employed clipped mean stacking to suppress artifacts and increase signal-to-noise \citep{2014PASP..126..158G}.
The pipeline produced template images on the fly for each CCD and pointing.
For images with no template coverage, the pipeline exited gracefully.
We are currently working to improve the template coverage of our pipeline by integrating more exposures that are publicly available from the NOAO archive.

\subsection{Photometric Calibration}
To photometrically calibrate our science images, we compared the magnitudes of stars extracted with \texttt{SExtractor} to the same stars on the reference images.
We then derived a zeropoint for the science images by taking the median zeropoint derived from each calibrator.
We also used this procedure to estimate the seeing on the science images, taking the median FWHM of each calibrator.
To choose calibrators, we selected only  objects with no \texttt{SExtractor} extraction error flags and a signal-to-noise ratio of at least 5.

\subsection{Image Subtraction, Source Identification, and Artifact Rejection}
For each pair of photometrically and astrometrically calibrated science images and templates, we used \texttt{scamp} to align the images to a common $x$-$y$ grid and the \texttt{HOTPANTS} \citep{2015ascl.soft04004B} implementation of the \cite{1998ApJ...503..325A} algorithm to convolve the images to a common PSF and perform a pixel-by-pixel subtraction.
We then ran \texttt{SExtractor} on the resulting difference images to identify sources of variability.
We rejected any objects that overlapped masked pixels on either the template or science images, had \texttt{SExtractor}\ extraction flags, had an axis ratio greater than 1.5, had a FWHM more than twice the seeing, had a PSF magnitude greater than 30, had a signal-to-noise ratio less than 5, or had a semi-major axis less than 1 pixel.
After making these initial cuts, we used the publicly available \texttt{autoScan}\ code \citep{2015AJ....150...82G}, based on the machine learning technique Random Forest, to probabilistically classify  the ``realness'' of the remaining extracted sources.
The code has been successfully used in past DECam searches for GW counterparts in independent difference imaging pipelines \citep[e.g.,][]{2017ApJ...848L..16S}.

\begin{figure*}
    \centering
    \includegraphics[width=1\textwidth]{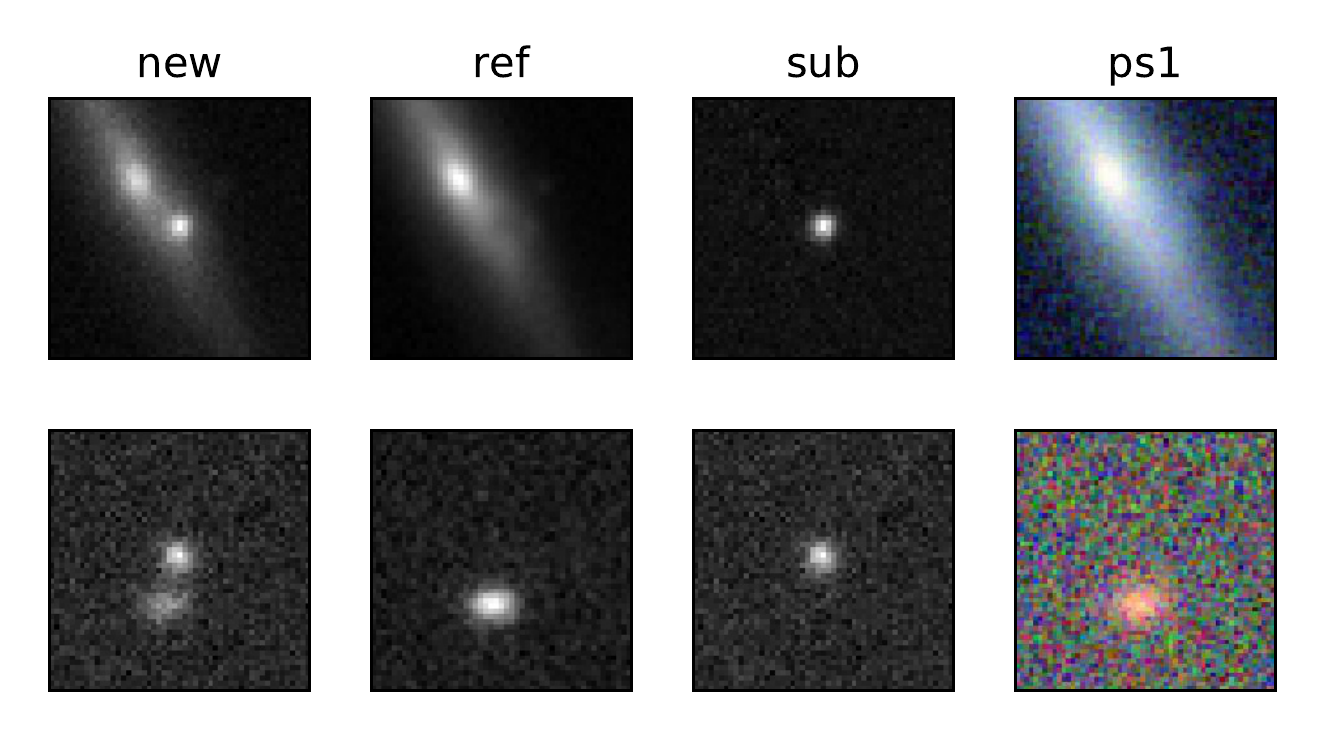}
    \caption{Postage-stamp cutouts of some counterpart candidates identified by the pipeline (top: DG19vkgf, bottom: DG19ytre).
    Each candidate has at least one detection in both $r$ and $z$, separated by at least 30 minutes (to reject asteroids).
    Full color images from Pan-STARRS1 (PS1) are shown for reference.
    }
    \label{fig:cutouts}
\end{figure*}

We pushed the candidates immediately and automatically to the GROWTH marshal, a dynamic web portal  for time-domain astronomy \citep{2019PASP..131c8003K}, where they were scanned by a team of roughly 10 scientists.
We reported nine promising counterpart candidates via GCN 2.5 hours before the end of our observations \citep{gcn5}.
We used the numerical score assigned by \texttt{autoScan} to each candidate to determine the order in which we looked at objects.
Using \texttt{autoScan} we were able to identify the transients we reported in the GCN by looking at less than 1\% of the candidate pool.
We also cross-matched each of our candidates against \textit{Gaia} DR2 \citep{2018A&A...616A...1G} to reject variable stars, the Minor Planet Center online checker\footnote{\url{https://minorplanetcenter.net/cgi-bin/checkmp.cgi}} to reject asteroids, and the Transient Name Server\footnote{\url{https://wis-tns.weizmann.ac.il/}} to reject known transients.
Figure \ref{fig:cutouts} shows images of two example candidates identified by the pipeline that were reported in the GCN, and Table \ref{tab:candidates} gives DECam photometry of  all candidates.

\subsection{Search Results and Pipeline Performance}
\begin{figure*}
    \centering
    \includegraphics[width=1\textwidth]{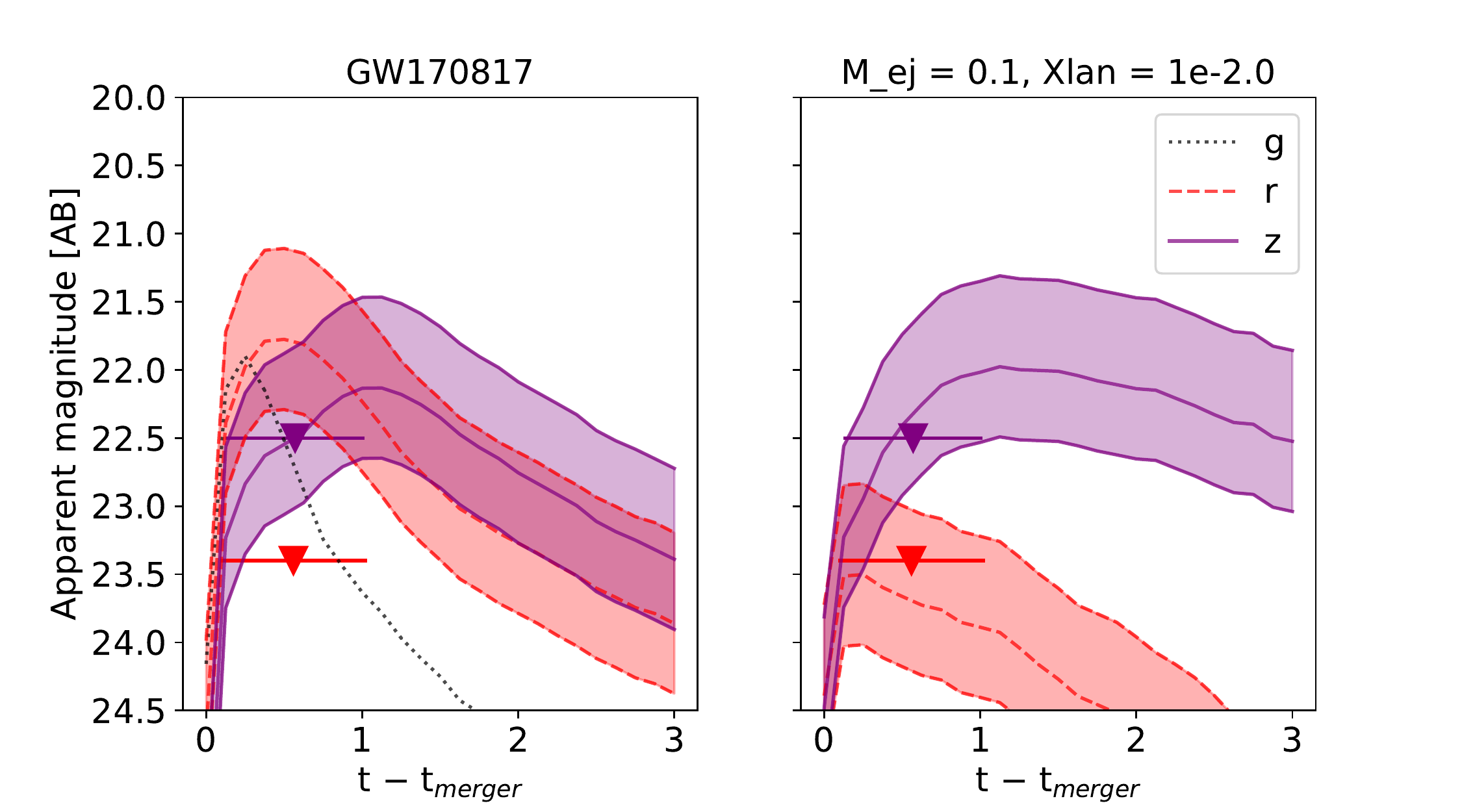}
    \caption{Kilonova models \citep{2017Natur.551...80K} at the distance range of S190426c. Markers indicate the 5$\sigma$ detection limit of our $r-z$ observations (not corrected for Galactic extinction). The left panel shows the light curve of the blue component of the BNS merger GW170817, the right panel presents a model with high ejecta mass (M$_{ej}$=0.1\,M$_{\odot}$) and large lanthanide content, a more plausible scenario in case of NS--BH merger.
    }
    \label{fig: models}
\end{figure*}

Processing each exposure with the pipeline required 16.7 minutes of wall-clock time, on average.
This fast turnaround time allowed us to detect transients quickly and rapidly communicate them to the community.
We identified 84,007 candidates: 45,587 in $r$-band and 48,931 in $z$-band images.
15,432 of our candidates had at least 2 detections.
 The measured depth reached during during our observations would have likely enabled the detection in both $r$ and $z$ bands of a GW170817-like event (Figure \ref{fig: models}, left panel).  
 Under the hypothesis that S190426c was in fact an NS-BH merger, the detection would have been more uncertain (Figure \ref{fig: models}, right panel) and longer exposure times would have aided the search.
One mildly red transient ($r-z = 0.3$ in DECam images), labelled DG19vkgf, was spectroscopically and photometrically followed-up by our team using the Hale 200-inch telescope (P200) at Palomar observatory  \citep{gcn6}.
A spectrum was obtained with the Double Beam Spectrograph \citep{1982PASP...94..586O} on P200.
Due to the high airmass and poor seeing conditions, the transient was not clearly identified in the trace, but the host redshift was confirmed to be $z = 0.04$ using the host emission lines.
Imaging with the Wafer Scale Imager for Prime (WASP) on P200 confirmed the presence of a point source at the transient location.

\cite{gcn9} followed up 2 events that we reported in \cite{gcn5}, DG19kplb and DG19ytre.  The authors performed photometric follow-up using the 1.5m telescope at the Observatorio de Sierra Nevada (Spain) starting on 2019-04-27 20:53 UT, spectoscopic follow-up using the 10.4m Gran Telescopio Canarias equipped with OSIRIS at La Palma (Spain) starting on 2019-04-27 21:40 UT.
Those observations allowed \cite{gcn9} to classify DG19kplb as a broad-line Type Ic supernova at redshift $z = 0.09123$ and DG19ytre as a Type Ia supernova at $z = 0.1386$.  The association of DG19kplb or DG19ytre with S190426c was therefore excluded.

When LIGO released a skymap  \citep{gcn7} that completely ruled out the possible association of DG19vkgf or any of the other transients we discovered using our pipeline with S190426c, we interrupted our photometric and spectroscopic follow-up of those sources.
We then focused our follow-up efforts on  transients discovered with northern hemisphere facilities that could access regions of higher localization probability (Kasliwal et al, in preparation; Bhalerao et al., in preparation).

\section{Conclusion}

 We carried out follow-up observations of the LIGO/Virgo gravitational wave trigger S190426c with DECam.
  Using an automated difference imaging pipeline, we were able to rapidly search our  data and publish candidates to the community before we completed our observations.
 Although we did not identify a counterpart with these observations, this enabled us to validate our DECam infrastructure for future events,  demonstrating that we can readily trigger, observe, scan, and detect transients on timescales, sky-areas and magnitude limits relevant for the discovery of gravitational wave counterparts.
   Availability of updated LVC sky maps on an even shorter timescale would allow us to more prudently use our telescope resources.
 In the future, we expect DECam to continue its important role as a discovery engine for gravitational wave counterparts.

 \begin{table*}[ht]
    \centering
    \begin{tabular}{ccc ccc c}
\hline \hline
Name & RA & Dec & Filter & $m$ & $\sigma_m$  & MJD \\
 & (J2000) & (J2000) &  &  &   &  \\

\hline
DG19ftnb & 167.595555 & $-$4.358792 & $z$ & 20.393 & 0.086 & 58599.99056 \\
&&& $r$ & 20.651 & 0.055 & 58599.96644\\
\hline DG19kqxe & 163.781705 & $-$0.237631 & $z$ & 21.059 & 0.117 & 58600.17142\\
&&& $r$ & 22.075 & 0.125 & 58600.13044\\
\hline DG19nmaf & 163.752355 & $-$1.486911 & $z$ & 21.603 & 0.102 & 58600.17142\\
&&& $r$ & 22.899 & 0.209 & 58600.13044\\
\hline DG19ouub & 171.473410 & $-$9.488396 & $z$ & 21.615 & 0.119&  58600.00142\\ && & $r$ & 22.123 & 0.102 & 58599.97506\\
\hline DG19vkgf & 165.844300 & $-$7.917442 & $z$ & 19.580 & 0.031 & 58600.19049\\
&&& $r$ & 19.888 & 0.017 & 58600.15045\\
\hline DG19zdwb & 167.296930 & $-$2.268391 & $z$ & 22.007 & 0.097 & 58599.99542\\
&&& $r$ & 22.803 & 0.117 & 58599.97024\\
\hline DG19zyaf & 163.471788 & $-$1.151129 & $z$ & 21.559 & 0.091 & 58600.17142\\
&&& r & 22.665 & 0.125 & 58600.13044\\
\hline\hline
DG19pklb & 168.658618 & -6.975466 & $z$ & 21.274 & 0.146 & 58599.99355 \\
&&& $r$ & 20.829 & 0.110 & 58599.96570\\
\hline DG19ytre & 167.760365 & 0.527199 & $z$ & 21.298 & 0.072 & 58600.11954\\
&&& $r$ & 20.693 & 0.040 & 58600.08185\\
\hline
    \end{tabular}
    \caption{Candidates discovered in real time using our transient detection pipeline and reported in \citep{gcn5}. The magnitudes ($m$) and uncertainties ($\sigma_m$) are in the AB system. The mid-point observing time is given in modified Julian days (MJD). \cite{gcn9} classified DG19kplb as a broad-line Type Ic supernova at redshift $z = 0.09123$ and DG19ytre as a Type Ia supernova at $z = 0.1386$, ruling out their association with S190426c. }
        \label{tab:candidates}
\end{table*}

\acknowledgments
D.A.G. and I.A. gratefully acknowledge Kathy Vivas, Steve Heathcote, and the NOAO staff for facilitating these target-of-opportunity observations.
This research used resources of the National Energy Research Scientific Computing Center (NERSC), a U.S. Department of Energy Office of Science User Facility operated under Contract No. DE-AC02-05CH11231.
D.A.G. acknowledges support from Hubble Fellowship grant HST-HF2-51408.001-A.
Support for Program number HST-HF2-51408.001-A is provided by NASA
through a grant from the Space Telescope Science Institute, which is operated by the Association of Universities for Research in Astronomy, Incorporated, under NASA contract NAS5-26555.
P.E.N. acknowledges support from the DOE through DE-FOA-0001088, Analytical Modeling for Extreme-Scale Computing Environments.
E.O.O. is grateful for support by  a grant from the Israeli Ministry of Science,  ISF, Minerva, BSF, BSF transformative program, and  the I-CORE Program of the Planning  and Budgeting Committee and The Israel Science Foundation (grant No 1829/12).
M.~W.~Coughlin is supported by the David and Ellen Lee Postdoctoral Fellowship at the California Institute of Technology. J.\,S.~Bloom, J.\,Martinez-Palomera, and K.\,Zhang are partially supported by a Gordon and Betty Moore Foundation Data-Driven Discovery grant.
J. Cooke is supported in part by the Australian Research Council Centre of Excellence for Gravitational Wave Discovery (OzGrav), CE170100004.

This work was supported by the GROWTH (Global Relay of Observatories Watching Transients Happen) project funded by the National Science Foundation under PIRE Grant No 1545949. GROWTH is a collaborative project among California Institute of Technology (USA), University of Maryland College Park (USA), University of Wisconsin Milwaukee (USA), Texas Tech University (USA), San Diego State University (USA), University of Washington (USA), Los Alamos National Laboratory (USA), Tokyo Institute of Technology (Japan), National Central University (Taiwan), Indian Institute of Astrophysics (India), Indian Institute of Technology Bombay (India), Weizmann Institute of Science (Israel), The Oskar Klein Centre at Stockholm University (Sweden), Humboldt University (Germany), Liverpool John Moores University (UK), University of Sydney (Australia) and Swinburne University of Technology (Australia).

This project used public archival data from the Dark Energy Survey (DES). Funding for the DES Projects has been provided by the U.S. Department of Energy, the U.S. National Science Foundation, the Ministry of Science and Education of Spain, the Science and Technology FacilitiesCouncil of the United Kingdom, the Higher Education Funding Council for England, the National Center for Supercomputing Applications at the University of Illinois at Urbana-Champaign, the Kavli Institute of Cosmological Physics at the University of Chicago, the Center for Cosmology and Astro-Particle Physics at the Ohio State University, the Mitchell Institute for Fundamental Physics and Astronomy at Texas A\&M University, Financiadora de Estudos e Projetos, Funda{\c c}{\~a}o Carlos Chagas Filho de Amparo {\`a} Pesquisa do Estado do Rio de Janeiro, Conselho Nacional de Desenvolvimento Cient{\'i}fico e Tecnol{\'o}gico and the Minist{\'e}rio da Ci{\^e}ncia, Tecnologia e Inova{\c c}{\~a}o, the Deutsche Forschungsgemeinschaft, and the Collaborating Institutions in the Dark Energy Survey.
The Collaborating Institutions are Argonne National Laboratory, the University of California at Santa Cruz, the University of Cambridge, Centro de Investigaciones Energ{\'e}ticas, Medioambientales y Tecnol{\'o}gicas-Madrid, the University of Chicago, University College London, the DES-Brazil Consortium, the University of Edinburgh, the Eidgen{\"o}ssische Technische Hochschule (ETH) Z{\"u}rich,  Fermi National Accelerator Laboratory, the University of Illinois at Urbana-Champaign, the Institut de Ci{\`e}ncies de l'Espai (IEEC/CSIC), the Institut de F{\'i}sica d'Altes Energies, Lawrence Berkeley National Laboratory, the Ludwig-Maximilians Universit{\"a}t M{\"u}nchen and the associated Excellence Cluster Universe, the University of Michigan, the National Optical Astronomy Observatory, the University of Nottingham, The Ohio State University, the OzDES Membership Consortium, the University of Pennsylvania, the University of Portsmouth, SLAC National Accelerator Laboratory, Stanford University, the University of Sussex, and Texas A\&M University.
Based in part on observations at Cerro Tololo Inter-American Observatory, National Optical Astronomy Observatory, which is operated by the Association of Universities for Research in Astronomy (AURA) under a cooperative agreement with the National Science Foundation.

The Legacy Surveys consist of three individual and complementary projects: the Dark Energy Camera Legacy Survey (DECaLS; NOAO Proposal ID \# 2014B-0404; PIs: David Schlegel and Arjun Dey), the Beijing-Arizona Sky Survey (BASS; NOAO Proposal ID \# 2015A-0801; PIs: Zhou Xu and Xiaohui Fan), and the Mayall z-band Legacy Survey (MzLS; NOAO Proposal ID \# 2016A-0453; PI: Arjun Dey). DECaLS, BASS and MzLS together include data obtained, respectively, at the Blanco telescope, Cerro Tololo Inter-American Observatory, National Optical Astronomy Observatory (NOAO); the Bok telescope, Steward Observatory, University of Arizona; and the Mayall telescope, Kitt Peak National Observatory, NOAO. The Legacy Surveys project is honored to be permitted to conduct astronomical research on Iolkam Du'ag (Kitt Peak), a mountain with particular significance to the Tohono O'odham Nation.

NOAO is operated by the Association of Universities for Research in Astronomy (AURA) under a cooperative agreement with the National Science Foundation.

\bibliography{ref}

\end{document}